\documentclass[]{spie}  

\usepackage{amsmath,amsfonts,amssymb}
\usepackage{graphicx}
\usepackage{eurosym}
\usepackage[colorlinks=true, allcolors=blue]{hyperref}

\title{CAMELOT - Concept study and early results for onboard data processing and GPS-based timestamping}

\author[a,b]{Andr{\'a}s P{\'a}l}
\author[a,b]{L{\'a}szl{\'o} M{\'e}sz{\'a}ros}
\author[c]{Norbert Tarcai}
\author[d,e,f]{Norbert Werner}
\author[d,g]{Jakub {\v R}{\'i}pa}
\author[f]{Masanori Ohno}
\author[f]{Kento Torigoe}
\author[f]{Koji Tanaka}
\author[f]{Nagomi Uchida}
\author[h]{G\'abor Galg\'oczi}
\author[f]{Yasushi Fukazawa}
\author[f]{Tsunefumi Mizuno}
\author[f]{Hiromitsu Takahashi}
\author[i]{Kazuhiro Nakazawa}
\author[c]{Zsolt V{\'a}rhegyi}
\author[j]{Teruaki Enoto}
\author[k]{Hirokazu Odaka}
\author[l]{Yuto Ichinohe}
\author[h,m]{Zsolt Frei}
\author[a]{L{\'a}szl{\'o} Kiss}

\affil[a]{Konkoly Observatory, MTA Research Centre for Astronomy and Earth Sciences, H-1121 Budapest, Konkoly Thege Mikl\'os \'ut 15-17, Hungary}
\affil[b]{Department of Astronomy, E\"otv\"os Lor\'and University, P\'azm\'any P. st. 1/A, Budapest, Hungary}
\affil[c]{C3S Electronics Development LLC., K{\"o}nyves K\'alm\'an krt. 12-14., Budapest, 1097, Hungary}
\affil[d]{MTA-E\"ot\"vos University Lend\"ulet Hot Universe Research Group, P\'azm\'any P\'eter s\'et\'any 1/A, Budapest, 1117, Hungary}
\affil[e]{Department of Theoretical Physics and Astrophysics, Faculty of Science, Masaryk University, Kotl\'a\v{r}sk\'a 2, Brno, 611 37, Czech Republic}
\affil[f]{School of Science, Hiroshima University, 1-3-1 Kagamiyama, Higashi-Hiroshima, Japan}
\affil[g]{Charles University, Faculty of Mathematics and Physics, Astronomical Institute, V Hole\v{s}ovi\v{c}k\'ach 2, 180 00 Prague 8, Czech Republic}
\affil[h]{Institute of Physics, E\"otv\"os University, P\'azm\'any P\'eter s\'et\'any 1/A, Budapest, 1117, Hungary}
\affil[i]{Department of Physics, Nagoya University, Furo-cho, Chikusa-ku, Nagoya, Aichi, Japan}
\affil[j]{The Hakubi Center for Advanced Research and Department of Astronomy, Kyoto University, Kyoto 606-8302, Japan}
\affil[k]{Department of Physics, University of Tokyo, 7-3-1 Hongo, Bunkyo, Tokyo 113-0033, Japan}
\affil[l]{Department of Physics, Rikkyo University, Nishi Ikebukuro 3-34-1, Toshimaku, Tokyo 171-8501, Japan}
\affil[m]{MTA-ELTE Astrophysics Research Group, P\'azm\'any P\'eter s\'et\'any 1/A, Budapest, 1117, Hungary}

\authorinfo{Further author information: A. P\'al: E-mail: apal@szofi.net}

\def\pinav{PiNAV-L1}
 
\begin{document} 
\maketitle

\begin{abstract}
Due to recent advances in nanosatellite technology, it is now feasible to 
integrate scintillators with an effective area of hundreds of 
square-centimeters on a single three-unit cubesat. We present the 
early test results for the digital payload electronics developed for the 
proposed CAMELOT (Cubesats Applied for MEasuring and LOcalising Transients) 
mission. CAMELOT is a fleet of nanosatellites intended to do full-sky monitoring 
and perform accurate timing-based localization of gamma-ray transients. 
Here we present
the early results on the GPS timestamping capabilities of the 
CAMELOT payload electronics, concluding that the investigated 
setup is capable to timestamp the received gamma-ray photons with an 
accuracy and precision better than 0.02 millisecond, which corresponds 
to a timing based localization accuracy of $\sim 3.5^{\prime}$. 
Further refinements will likely allow us to improve 
the timing accuracy down to the sub-microsecond level. 
\end{abstract}



\section{Introduction}

The rapid localization and characterization of the high energy phenomena associated with gamma-ray bursts (GRBs) are among the most important challenges of today's observational astrophysics. The recent simultaneous detection of a GRB and the gravitational waves source GW170817\cite{gw2016a,gw2016b,gw2017a} further increased the interest in the rapid localization of the electromagnetic counerparts of GW sources. One of the many possible ways of localizing GRBs is 
based on their precise timing using numerous 
space-borne scintillator-based soft gamma-ray detectors evenly distributed in the low Earth orbit. Here, and in two companion papers submitted to this conference, we propose the CAMELOT  
(Cubesats Applied for MEasuring and LOcalising Transients)  mission\cite{werner2018}, which is a fleet of nanosatellites of the 3U cubesat standard equipped with such sensitive detectors\cite{ohno2018}. Recently, we 
demonstrated\cite{torigoe2018,ohno2018}
that thin caesium iodide (CsI) crystals equipped with multi-pixel photon counters 
(MPPCs) can be integrated on 3U cubesats and used to detect and characterize GRBs.

Our nanosatellite design includes four separate CsI 
scintillation detectors on two perpendicular sides of the satellites. Each of the scintillator crystals is equipped 
with two separate MPPC sensors\cite{werner2018}. 
This setup allows us to perform a readout supported 
by coincidence detection, and the detector geometry provides 
additional constraints on the localization, independently of timing. 
Thereore, to effectively achieve our scientific goals and in the same time provide redundancy, eight separate channels are needed to be monitored and analyzed simultaneously.

Due to the limited power, redundancy, reprogramming and reconfiguration 
requirements, heat production and the onboard space availability for the 
circuits, a careful trade-off analysis is needed to choose the most suitable components and configuration for the payload electronics. We study the use of field-programmable gate arrays and microcontroller units, including signal shaping, triggering, time-stamping as well as 
housekeeping, brownout and watchdog functionality. In addition, several 
separate components are also needed to be included within the payload 
electronics, most importantly a GPS receiver. The recent advancement of 
such receivers allows us to perform onboard time-keeping and time-stamping 
of GRB events with a sub-millisecond accuracy. 
Assuming low-Earth orbit and a timing accuracy of the order of tens of microseconds, the 
corresponding celestial position accuracy will be in the range of 
10 arcminutes, which is well matched for rapid 
ground-based or space-borne follow-up campaigns\cite{ohno2018}.


In this paper we summarize the key concepts of a GPS timestamping
unit designed for precise and accurate retrieval of instances related to
scintillation events inducted by gamma-ray photons. This unit is based on the 
\pinav{} GPS receiver which provides a NMEA-compatible UART stream and 
has a Valid Position Pulse (VPP) output for which a timestamp is associated with a
$100\,{\rm ns}$ resolution and better-than $1\,{\rm\mu s}$ accuracy
(within 2-$\sigma$). Like in the case of many commercial GPS receivers,
the \pinav{} model provides both the NMEA sequences and the VPP output with
a period of 1 seconds.

In order to make laboratory testing easier, we interfaced this 
unit with an onboard USB and Ethernet interface. The former one 
acts as the primary power supply of this GPS test unit as well as provides
an RS485 interface for the host microcontroller (via an FT232 USB-UART bridge). 
The Ethernet subsystem
is based on a 100Mbit, full-duplex and fully integrated TCP/UDP/IP/MAC/PHY core
and allows a low-latency communication interface in order to compare 
the timestamping precision and accuracy with Network Time Protocol (NTP)
and/or another UDP/IP based services. 

\begin{figure}
\begin{center}
\resizebox{160mm}{!}{\includegraphics{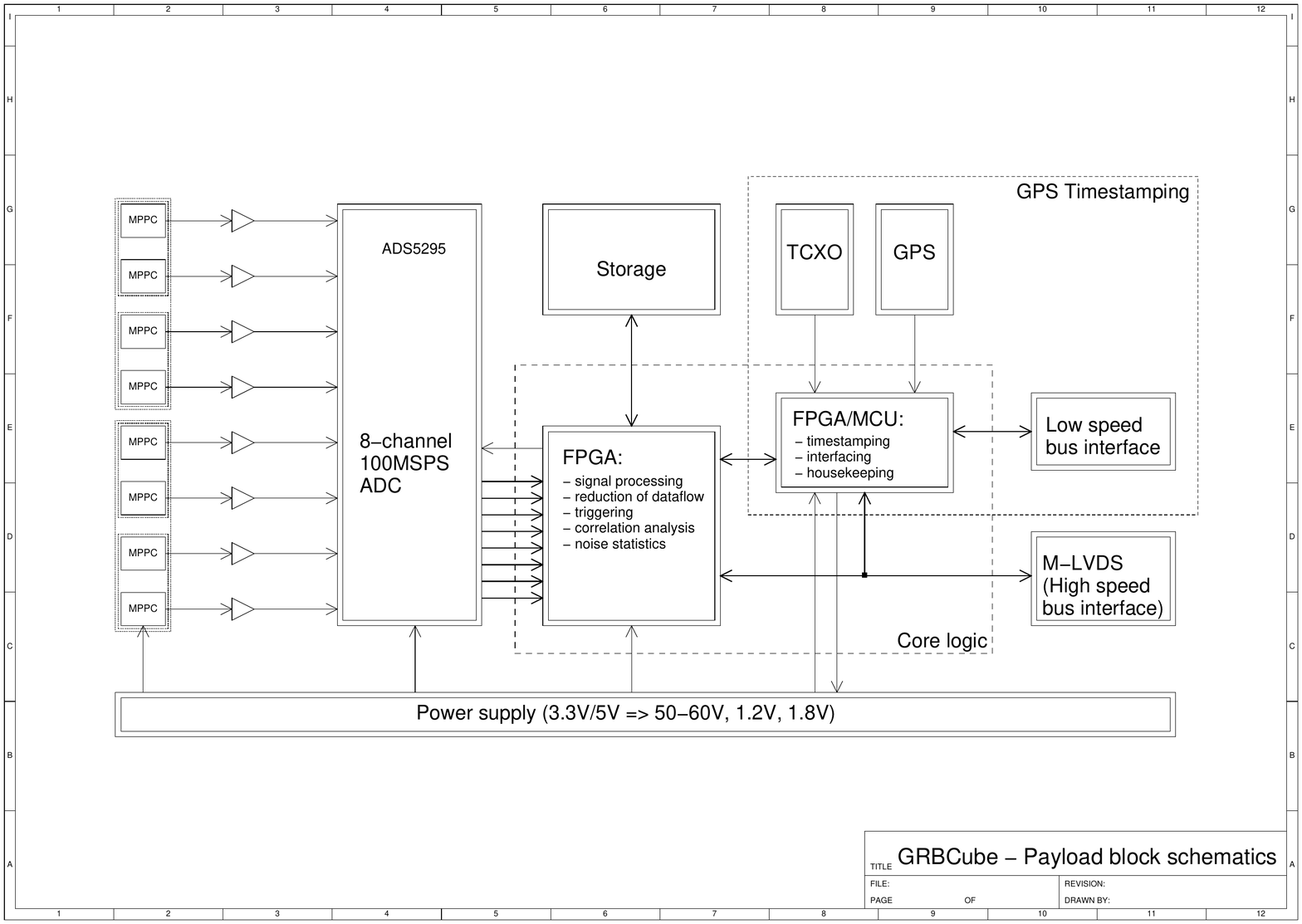}}%
\end{center}\vspace*{-5mm}
\caption{Simplified block diagram of the CAMELOT payload. The dashed box marks
the ``core logic'' of the payload while the dotted box marks the logic
relevant to the GPS timestamping unit. As it can be seen from this diagram, the 
logic of the GPS timestamping can be merged with the core 
logic, i.e. the implementation of all of the necessary features -- mainly the digital signal processing and
the timestamping -- can be achieved in a single controller.}
\label{fig:camelotpayload}
\end{figure}

\section{Theory of operations}

The GPS timestamping unit is an essential part of the digital core logic
of the CAMELOT payload. As it can be seen in Fig.~\ref{fig:camelotpayload},
the functionalities of the timestamping and the digital signal processing
are highly overlapping. Therefore, in the final (flight) version of the 
payload, we have to consider possibilities of merging these logics and/or
the use of separate boards.

\subsection{Features}

\begin{figure}
\begin{center}
\resizebox{160mm}{!}{\includegraphics{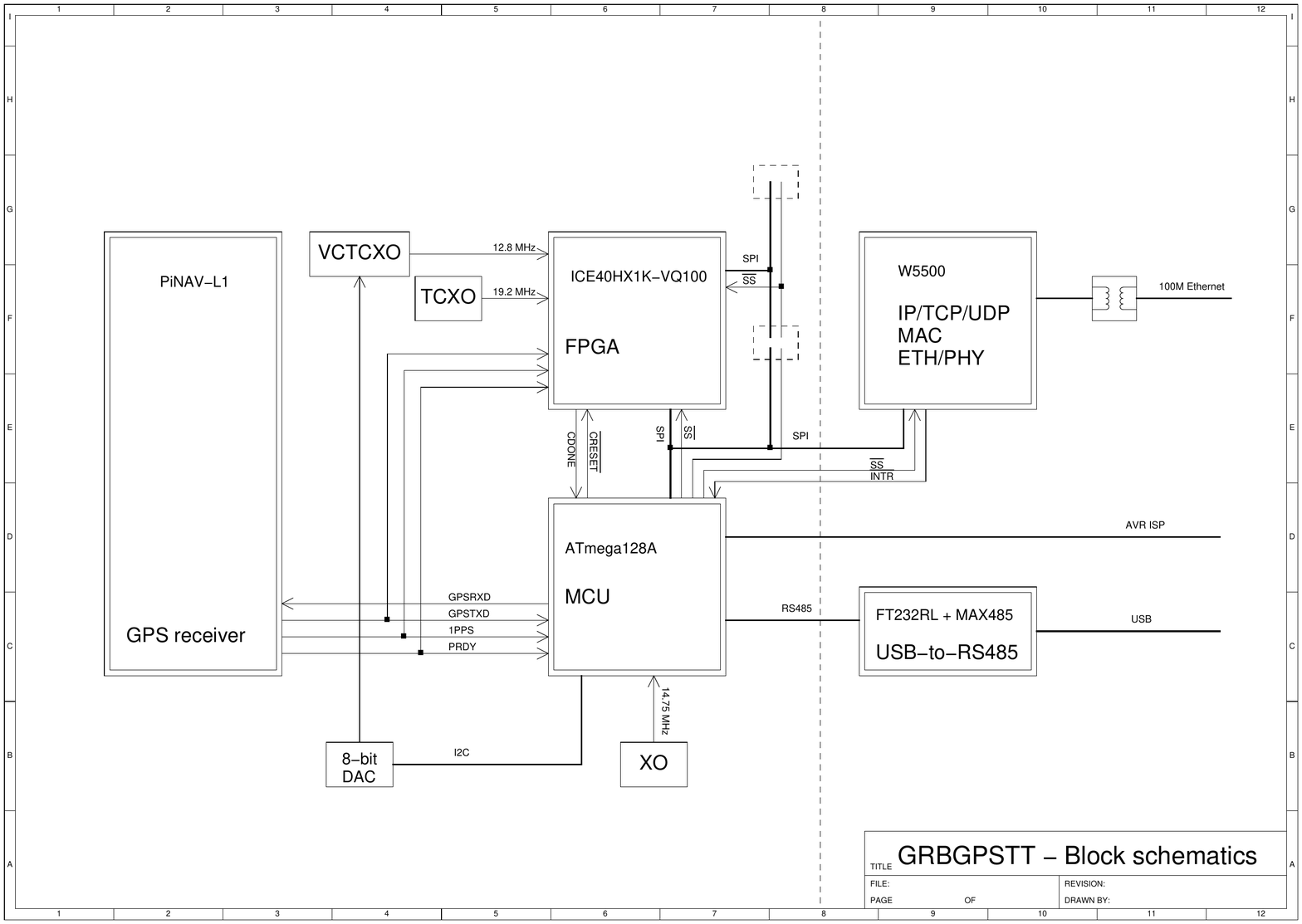}}%
\end{center}\vspace*{-5mm}
\caption{Block diagram of the CAMELOT GPS test board.
The dashed line separates the parts which are essential for the actual 
timestamping (the left part, i.e. where the equivalent functionality 
is needed to be implemented
on a flight model as well) and the blocks used for ground-based testing
(most importantly, USB and Ethernet downlink connectors and physical interfaces).}
\label{fig:ggtt-block}
\end{figure}

In order to test the GPS characteristics of the indented 
receiver, we designed a board (GRB GPS test board, GGTT-board or GGTT 
for short) intended to provide the full functionality of the timestamping 
logic (see also the upper-right dotted box of Fig.~\ref{fig:camelotpayload}).
In the following, we summarize the most essential components of the 
GPS timestamping unit as well as the GGTT board. We refer here to 
Fig.~\ref{fig:ggtt-block}, showing the block diagram of the circuit.
\begin{itemize}
\item The timestamping core is hosted by the FPGA which acts an SPI 
slave device. The timestamps are acquired during the falling edge
of the \#SS (slave select) line, i.e. in the \#SS: $1\to0$ transitions.
\item The corresponding timestamp is clocked out via the MISO line.
\item The MCU parses the ASCII NMEA sequences, converts them to binary
and uploads them to the FPGA every second.
\item The MCU acts as a tester unit, i.e. capable to ask the FPGA for
a timestamp and in parallel, ask a remote NTP server for a timestamp. 
\item The board hosts two precision temperature compensated crystal
oscillators (TCXO's) where one of these TCXOs can also be tuned (i.e. it is
a voltage controlled temperature compensated crystal oscillator, VCTCXO).
These oscillators are used as frequency standards between the 
subsequent VPPs and/or during the loss of GPS signal. 
\item The high-latency USB interface can be used to monitor the board.
The MCU itself is interfaced using an UART-RS485 physical interface
while the USB UART stream is also converted to RS485 signal levels.
This dummy RS485 line will be used to simulate the satellite bus. 
\end{itemize}
We note here that the third point above, i.e. the parsing of the NMEA
sequences by the MCU and upload the binary timestamps to the FPGA
does not decrease the accuracy and/or the precision of the unit at all.
This kind of implementation just makes the development and the 
testing phase easier. 
In addition, this can also be considered as a temporal
feature since the FPGA itself can be programmed to parse the NMEA stream. 

\begin{figure}
\begin{center}
\resizebox{120mm}{!}{\includegraphics{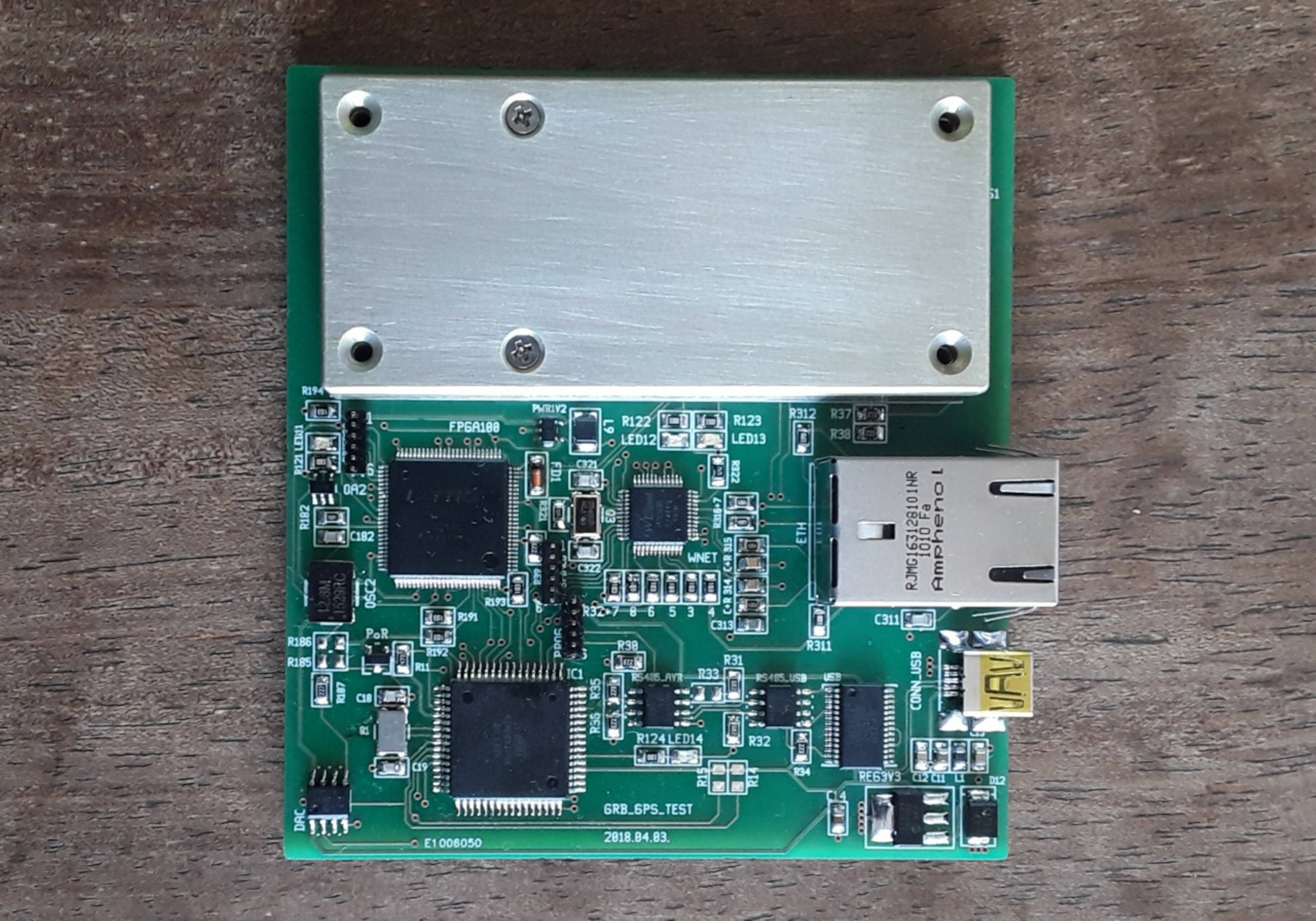}}%
\end{center}
\caption{A photo of the CAMELOT GPS test board. The connectors (Ethernet and
mini-USB) are located on the lower-right corner. The board itself is dominated
by the engineering model of the \pinav{} GPS receiver which has the same
physical form factor and electrical characteristics as the flight model.
The size of the board is $3380\times3660$\,mm.}
\label{fig:ggtt-board}
\end{figure}

The GRB GPS test board is shown in Fig.~\ref{fig:ggtt-board}. The actual size of the board 
is $86\times93\,{\rm mm}$, which corresponds to the cubesat 
standards. See the figure caption for more details. 

\subsection{Interfacing}

\begin{figure}
\begin{center}
\resizebox{160mm}{!}{\includegraphics{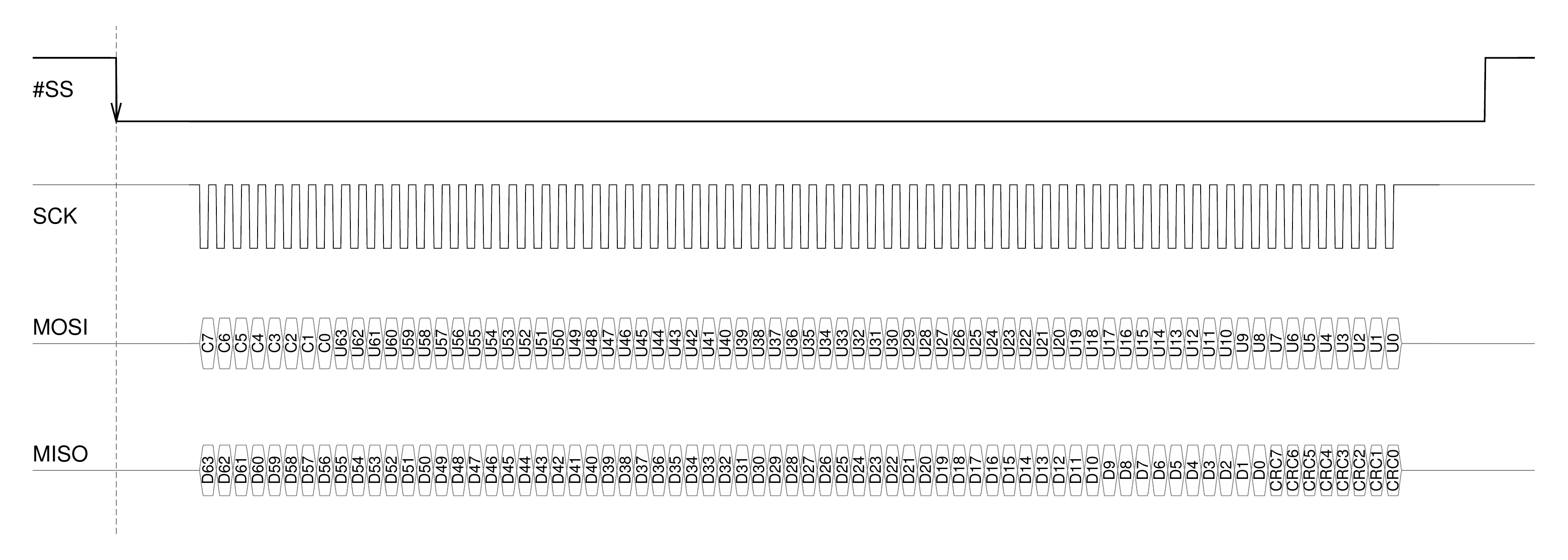}}%
\end{center}\vspace*{-5mm}
\caption{The Serial Peripheral Interface (SPI) communication as implemented
in the timestamping module. This implementation follows the SPI mode 3 (CPOL=1,
CPHA=1) standard, hence, the SPI master must be configured according to
this mode. In the MOSI line, the master must issue a single-byte command,
followed by a 8-byte long optional parameter set. In the MISO line,
the slave clocks the actual timestamp, followed by an 8-bit CRC.
Currently, two commands are implemented. The command
\texttt{C[7:0]=0b00000000} returns the timestamp corresponding to
the falling edge of the slave select (\#SS) line. 
In addition, the command \texttt{C[7:0]=0b01010101=0x55} sets the 
reference timestamp
to \texttt{U[63:0]} for the previous rising edge of the GPS 1PPS signal.
The timestamps are always considered in TAI in the units of UNIX time. 
The upper part, \texttt{D[63:32]} are for seconds, while the lower part
\texttt{D[31:0]} is for fractions of the second.
In other words, \texttt{D[63:0]} is a fixed-point floating number 
with a resolution of $2^{-32}$ seconds. In the current implementation,
the actual resolution is close to $1\,{\rm \mu s}$, i.e. the lower
12 bits are zero: \texttt{D[11:0]=0b000000000000}. 
We note that the format is exactly the same as it is used in the 
Network Time Protocol (besides the zero-point, which is 1900-01-01 in 
the case of NTP and 1970-01-01 here).
The 8-bit CRC output of \texttt{CRC[7:0]} is computed using the CRC-8-CCITT
standard, i.e. using the polynomial $x^8+x^2+x^1+1$. }
\label{fig:spidiagram}
\end{figure}

The core of the GPS timestamping unit is implemented in an FPGA hardware
where the corresponding IP core can be directly wired to the I/O ports of the 
chip and can be re-used in more complex FPGA-hosted firmwares.
The primary interface signaling, i.e. the SPI slave operations
(chip select, clocking, data lines, etc.) can be seen in 
Fig.~\ref{fig:spidiagram}. 

\section{Test results}

We divide the following section into two parts. First, we discuss the results
of the timestamping precision compared with NTP servers while in the second
part, we compare the accuracy of the receiver timestamps and algorithm
with an independent off-the-shelf low-resolution timestamping module
provided by Finger Lakes Instrumentation (FLI). This latter module was
designed to perform timestamping of CCD images (beginning and ending 
instances of exposures, open and close times of shutters, etc.)
in astronomical imaging applications, where accurate timing
information is needed\cite{hirt2014}. 

\subsection{Comparison of the precision with NTP-based reference clocks}

The testing of the GGTT device was conducted as follows: an NTP query
is issued via the integrated UDP/IP/MAC/PHY stack and the FPGA is
asked to perform a timestamping just before sending the appropriate
NTP query UDP packet and right after receiving the NTP answer packet. 
Let us denote these
timestamps by $T_1$ and $T_2$ while the returned NTP timestamp is
denoted by $T$. The precision of the process can be characterized by the
scatter of the $\Delta T=T-(T_1+T_2)/2$ value while the overall accuracy is 
limited by the long-term trends. The process needs in total $T_2-T_1$ time.

\begin{figure}
\begin{center}
\resizebox{80mm}{!}{\includegraphics{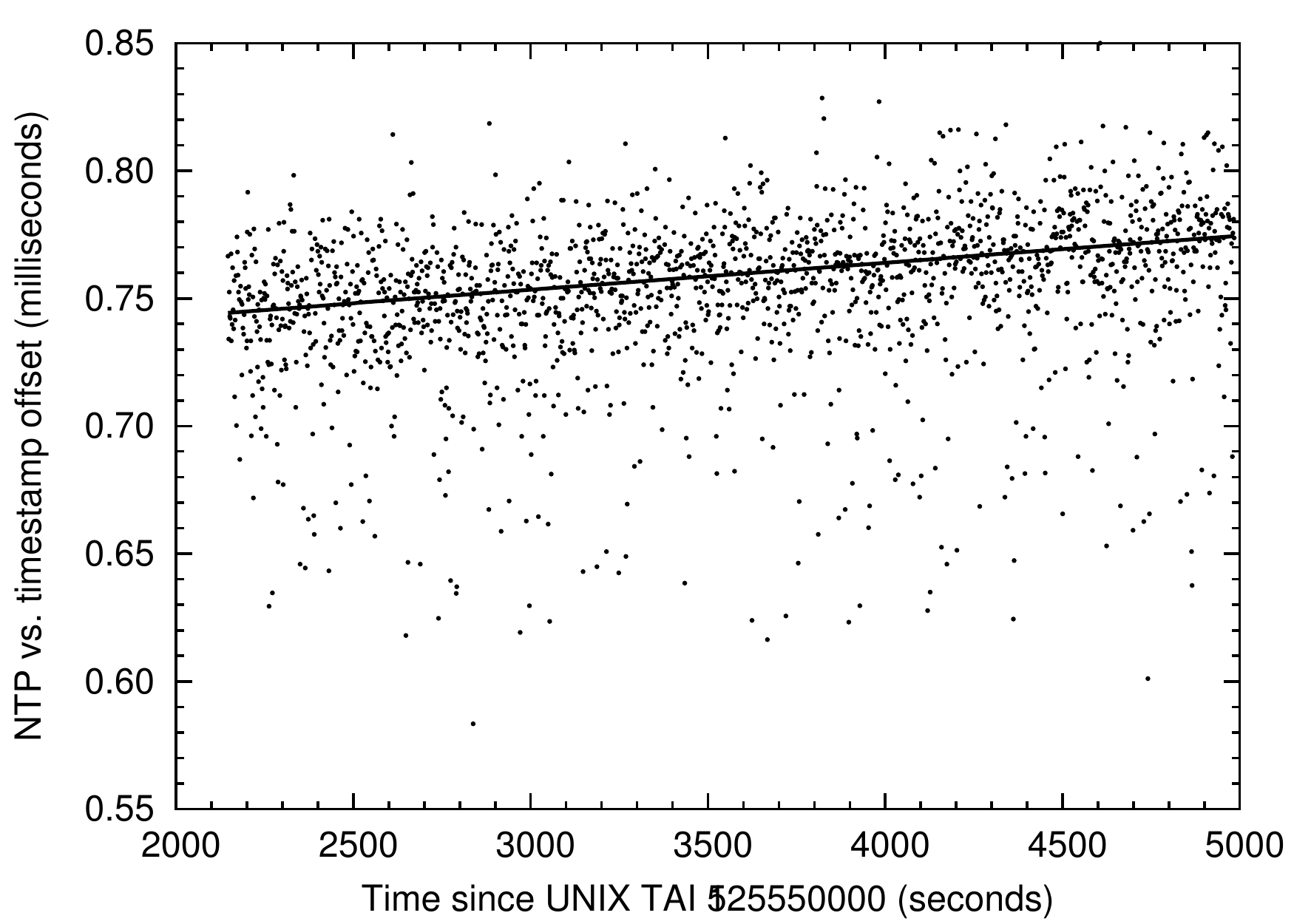}}%
\resizebox{80mm}{!}{\includegraphics{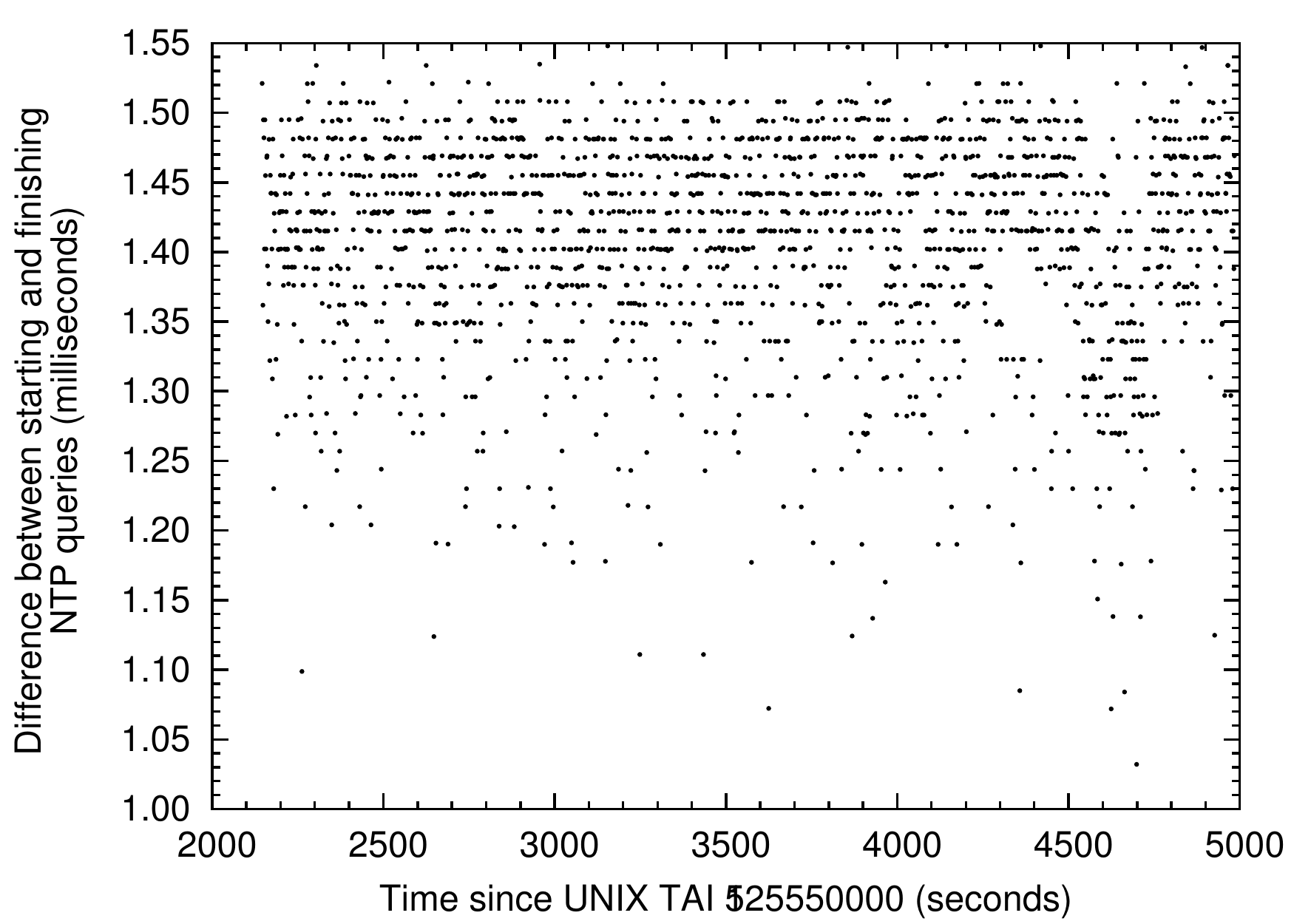}}%
\end{center}
\caption{{\it Left panel:} The difference between the NTP timestamp
and the mean of the GPS timestamps before and after the NTP query.
The scatter of the plot (i.e. the overall precision) is 19 microseconds
while the systematic offset (with a slight long-term drift) is 
$\sim 0.75$ milliseconds. {\it Right panel:} The $T_2-T_1$ time, i.e.
the time needed for the NTP query during the test run. 
The horizontal structure is due to the quantized polling of the UDP/IP/MAC/PHY
interface. The overall scatter of this quantity is $\sim 75$ microseconds, far
more than the $\sim 13$ microsecond polling period.}
\label{fig:stamp1}
\end{figure}

\begin{figure}
\begin{center}
\resizebox{80mm}{!}{\includegraphics{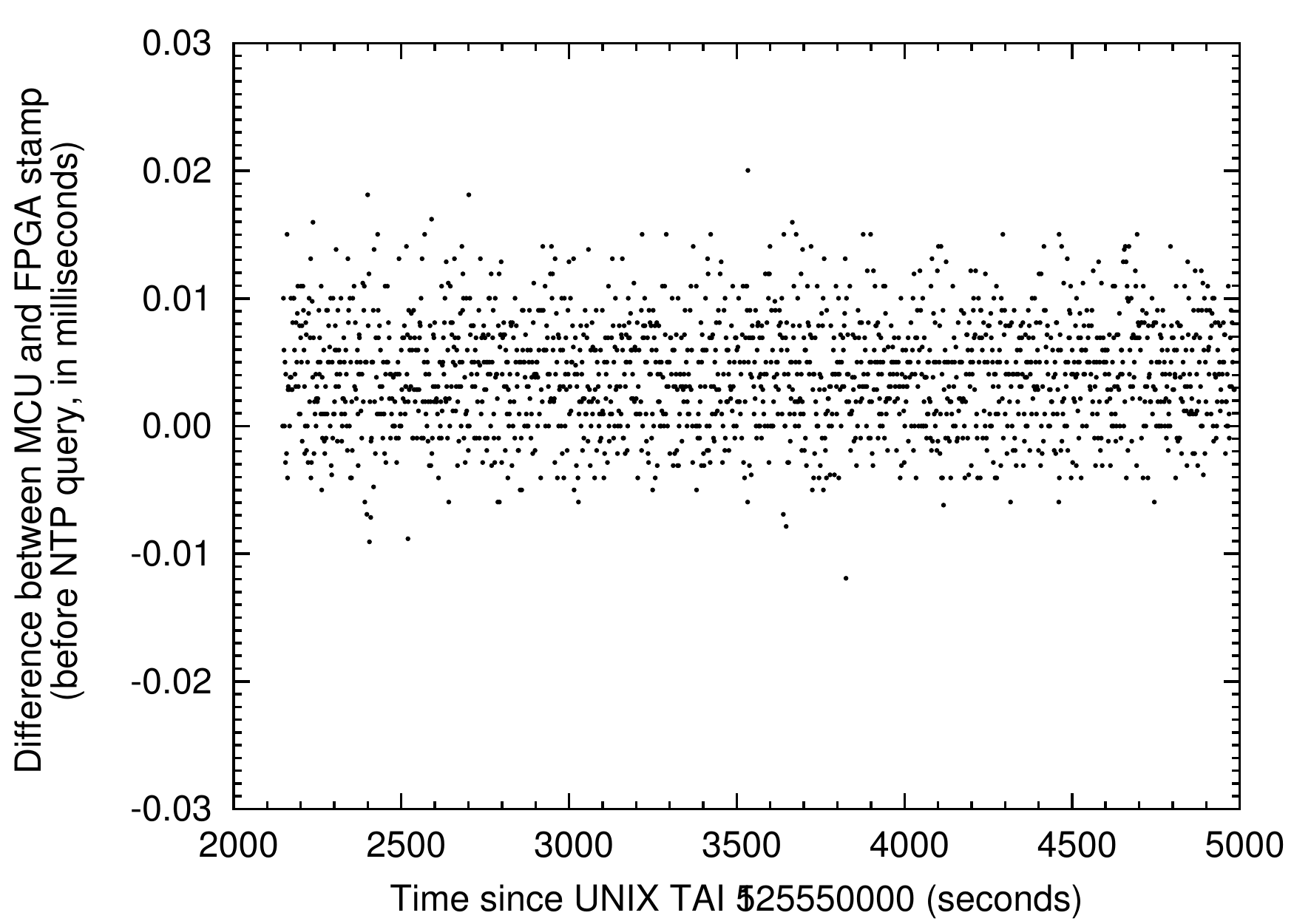}}%
\resizebox{80mm}{!}{\includegraphics{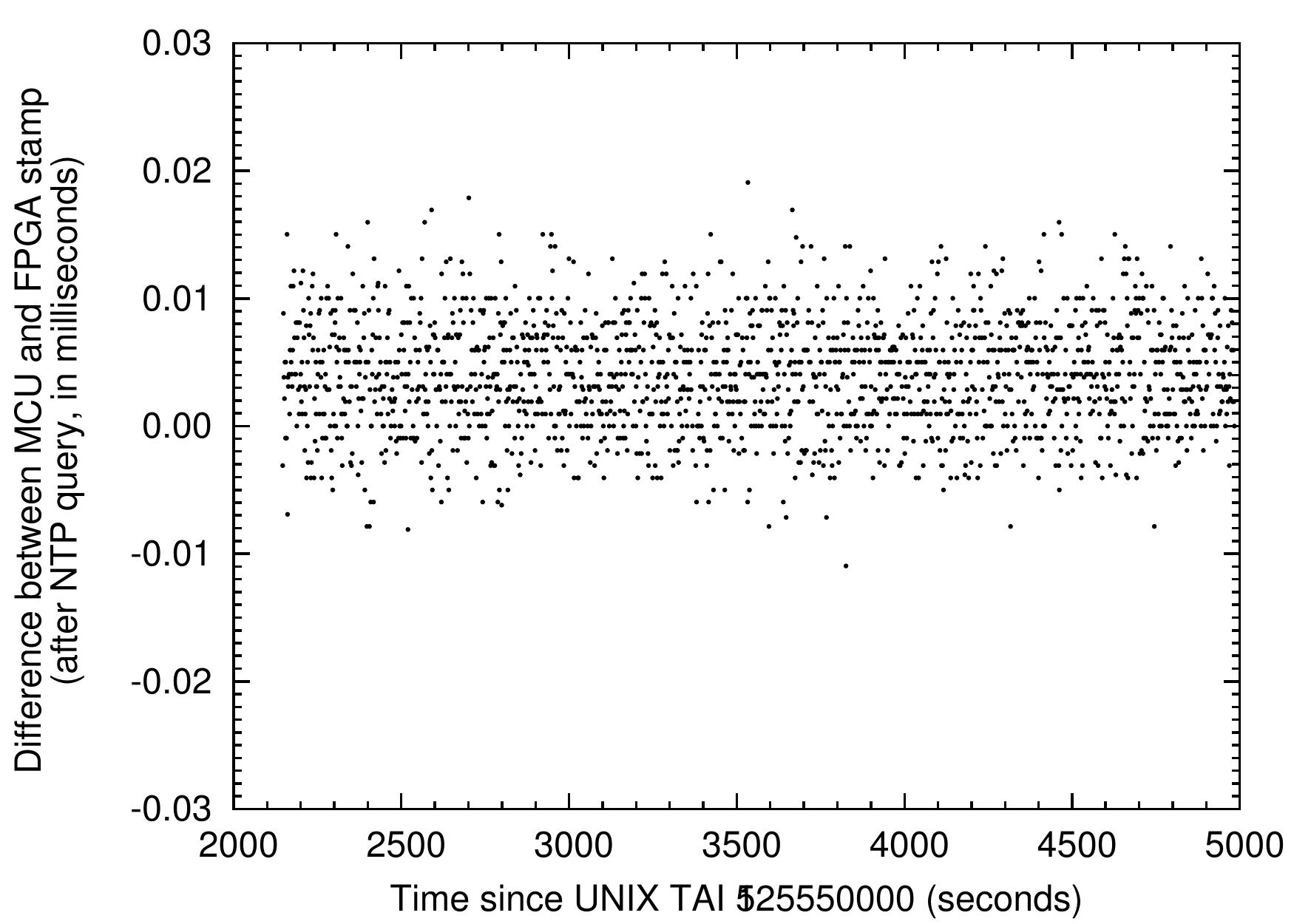}}%
\end{center}
\caption{The differences between the MCU-based and the FPGA-based
timestamps before ({\it left panel}) and after ({\it right panel})
the execution of the NTP queries. The scatter
of both plots is around 4.3 microseconds. The slight offset of 3.9
microseconds between the two values are simply due to the relatively 
lengthy overflow-bit checking 
procedure of the process related to the timer counter sampling. 
The scatter itself is due to the fact that the rising edge of the GPS
VPP output is sampled in a loop instead of triggering an input capture
event and/or input change interrupt.}
\label{fig:stamp2}
\end{figure}

In Fig.~\ref{fig:stamp1} we plotted the values of $\Delta T$ and $T_2-T_1$
during an 50-min long test series where the NTP queries and the corresponding
timestamps have been executed subsequently with a random delay between 1
and 2 seconds. For the tests, we used an off-site Stratum 1 NTP server, 
located physically a few kilometers away from the test site and within 
a topological distance of 6 hops\footnote{We note here that the hop
distance does not take into account the number of switching elements.
In our experiment, there were 3 additional
network switches in the local area network, i.e. between the GGTT board
and the core uplink router.}.
In Fig.~\ref{fig:stamp2} we show the differences between a very crude MCU-based
implementation of the timestamping as well as the FPGA-based timestamping
algorithm. 

We note here that the systematics offset of $\sim 0.75$ milliseconds in the 
$\Delta T$ values are well below the intended accuracy of NTP -- 
which states
a $\sim 10$ millisecond limit in the case of off-site servers and a few milliseconds
in the case of on-site, local area network (LAN) servers. 
In the near future, it is worth to repeat the experiment involving an 
off-the-shelf Stratum 1 server on the same LAN. 

\subsection{Comparison of the accuracy with low-precision GPS timestamp units}

We also performed a comparison of our timestamping algorithm with
an independent, low-precision GPS timestamping unit provided by 
Finger Lakes Instrumentation. This GPS module of FLI accepts LVTTL input 
signals and is designed to capture two time events, corresponding to CCD image
exposures and/or shutter open-close instances. Namely, the first timestamp is
caught during the rising edge of the input signal while the second timestamp
is caught during the falling edge of the input signal. We used this
dual feature to capture the instances corresponding to the send and receive
times of the NTP/UDP packets. 

\begin{figure}
\begin{center}
\resizebox{160mm}{!}{\includegraphics{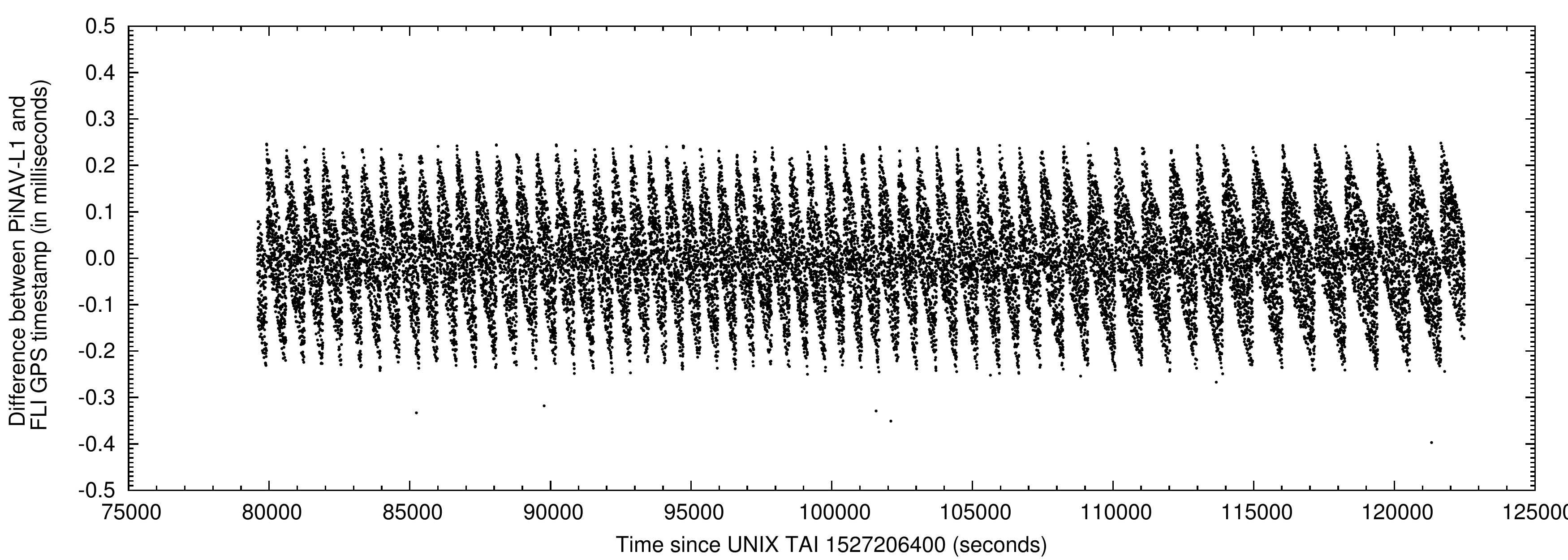}}
\end{center}
\caption{The difference between the \pinav{} timestamping and the FLI GPS 
timestamp unit for a longer run of $\sim 12$ hours. Although the resolution of 
the latter module is limited to $0.25$ milliseconds and its local oscillator
has a frequency offset of $0.2-0.3$ ppm (depending on the ambient
temperature), its accuracy compared with the \pinav{} output is 
within $2-3$ microseconds.} 
\label{fig:stampfligps}
\end{figure}

This module provided by FLI has a resolution limited to 12 bits, i.e. 
its effective resolution is $1/4000$ seconds. However, besides this 
precision and the measured level of frequency offsets of its internal
oscillators, we found that the accuracy of the \pinav{} module and the 
FLI GPS unit is within a few microseconds. In addition, considering the 
systematic drifts due to the limited precision and interpreting the residuals
as pure statistical noise, the precision is in the range of $0.1$ milliseconds
(which coincides with the standard deviation of the $[-1/8000,+1/8000]$
interval implied by the effective resolution of this module). 
See also Fig.~\ref{fig:stampfligps} for more details. 

\section{Trade-offs}

Even in this simple test board system, several questions arise regarding to
various trade-offs of a future flight model of the board. These problems
are related to the implementation, redundancy and interfacing. In the
following, we summarize the main points of these trade-offs. 
\begin{itemize}
\item {\bf FPGA vs. MCU.} In the current implementation, timestamping can
be implemented both by the FPGA and the MCU. While the theoretical accuracy
and precision of the timestamping is in the range of the FPGA master clocks
(in the order of few tens of MHz), the accuracy of the MCU is much lower.
While a complex MCU (such as the ATmega128A used in our experiments)
hosts interfaces and features such as an Input Capture Pin, the
overhead associated with the parsing of the corresponding registers yields
a much slower response and many additional circumspections (for instance,
handling the overflows in the timer counters, etc.)

\item {\bf MCU type and architecture.} In the current implementation,
an 8-bit AVR ATmega128A controller is used to acts as the "main computer"
of this board. It is worth to investigate the accommodation of
other types of MCUs, likely 32-bit ARM processors for this purpose. 
The MCU should host at least two independent UARTs in order to 
interface both with the GPS receiver and the satellite bus. The minimal
peripheral requirements for the MCU are hence:
\begin{itemize}
\item 2 UARTs;
\item 1 SPI master;
\item 1 input capture line and/or 1 input change interrupt line.
\end{itemize}
Of course, test models like this unit might be needed to feature
additional interfaces, such as Ethernet or CAN. Besides RS485, CAN bus 
can also be considered as the primary interface to the satellite platform.

\item {\bf FPGA bitstream and configuration.} 
In the current implementation of our GGTT board, the FPGA 
bitstream is stored in the flash memory of the microcontroller, allowing 
to upload the bitstream of in slave or peripheral mode 
(as viewed from the side of the FPGA). While this setup eliminates the need of 
an additional mission-critical part (i.e. a serial flash memory storing
the FPGA bitstream) as well as makes the interfacing
and the remote firmware upgrade easier (i.e. there is no need for a
multi-master SPI wiring), the cons include the fact that both 
the MCU and the FPGA needs to operate in a seamless manner.
It is worth to investigate whether a simple multiplexer (possibly implemented 
in the form of open-drain logic on the MISO/MOSI lines with stand-alone 
transistors) can be inserted in the circuit in order to allow 
both peripheral-mode and processor-mode configuration of the FPGA CRAM. 

\item {\bf Congestion.}
While the time stamping information can be clocked out quite fast
(within a few microseconds), it might be an issue that two overlapping
events occur within the timeframe of the SPI operations. This is 
unlikely due to the nature of the scintillator detections and the
expected gamma photon rate but the full system needs to handle it 
at some level (e.g. associate the same timestamp for more events but
count the scintillation events properly). This issue is beyond the 
capabilities of a simple SPI interface. 
\end{itemize}

\section{Conclusions}

As a conclusion, we confirm that the investigated setup is capable to 
timestamp the received gamma-ray photons with an accuracy and precision better than $\sim 0.02$ milliseconds. This value corresponds to a localization accuracy of $10^{-3}$ radians, which is equivalent to $\sim 3.5^{\prime}$. 
We note here that additional refinements and further tests (by involving
LAN-hosted Stratum 1 NTP servers and/or more precise independent 
timestamping units) will likely allow us to improve the timing accuracy down
to the sub-microsecond level. Such improvements will allow us to measure 
the positions of bright sources with a sub-arcminute accuracy. However,
we have to keep in mind that the timing accuracy also depends on 
the number of photons received, i.e. photon statistics -- which does not 
allow sub-arcminute accuracy in the case of fainter GRBs. 


\acknowledgments 
This work was supported by the Lend\"ulet LP2016-11 and LP2012-31 
grants awarded by the Hungarian Academy
of Sciences, as well as by the grant GINOP-2.3.2-15-2016-00033.
This work was also supported by Hiroshima University, 
JSPS KAKENHI Grant Number 17H06362.
In this project, we involved many 
free \& open source software packages, including 
gEDA (for schematics and PCB design), 
AVR-GCC (for AVR MCU programming), 
the GCC ARM Embedded toolchain (for ARM Cortex-M MCU programming)
and the Yosys Open Synthesis Suite with additional
tools provided by Project IceStorm (including Arachne-PNR and IcePack). 
We also thank Finger Lakes Instrumentation for the support for the low-level
programming of their GPS timestamping module. 


\bibliography{ggtt-v03} 
\bibliographystyle{spiebib} 


\end{document}